\documentstyle[aps,prl,multicol,psfig]{revtex}

\begin{document}

\title{\vbox to 0pt {\vskip -1cm \rlap{\hbox to \textwidth {\rm{\small
SUBMITTED TO PHYS. REV. LETT.\hfill}}}}Statistics of
Dissipation and Enstrophy Induced by a Set of Burgers Vortices}

\author{Guowei He$^{1,4}$, Shiyi Chen$^{1,2}$,
Robert. H. Kraichnan$^{3}$, Raoyang Zhang$^{1}$, 
and Ye Zhou$^{2}$ }

\address{${}^{1}$Theoretical Division and Center for Nonlinear Studies,
Los Alamos National Laboratory, Los Alamos, NM 87545\\
${}^{2}$IBM Research Division, T. J. Watson Research Center,
P.O. Box 218, Yorktown Heights, NY 10598 \\
${}^{3}$396 Montezuma 108, Santa Fe, New Mexico, 87501-2626, \\
${}^{4}$ LNM, Institute of Mechanics, Chinese Academy of Sciences, 
Beijing, 100080, P. R. China}

\maketitle

\begin{abstract}Dissipation and enstropy statistics are calculated for an
ensemble of modified Burgers vortices in equilibrium under uniform straining.
Different best-fit, finite-range scaling exponents are found for
locally-averaged dissipation and enstrophy, in agreement with existing
numerical simulations and experiments. However, the ratios of dissipation and
enstropy moments supported by axisymmetric vortices of any profile are finite.
Therefore the asymptotic scaling exponents for dissipation and enstrophy
induced by such vortices are equal in the limit of infinite Reynolds number.

\end{abstract}

\vskip 4mm

\begin{multicols}{2}

Recent direct numerical simulations (DNS) \cite{1} have revealed spatial
structures with regions of intense dissipation density $\epsilon =
\case{1}{2}\nu(\partial u_i/\partial x_j + \partial u_j/\partial x_i)^2$ and
enstrophy density $\Omega = \omega^2$, where $\nu$ is kinematic viscosity,
$\bbox{\omega} = \bbox{\nabla}\times{\bf u}$ is vorticity, and ${\bf u}$ is
the fluid velocity. The spatial distributions of enstrophy density and
dissipation density are qualitatively different. Intense vorticity has a
tube-like (or filament-like) structure, while intense dissipation typically
surrounds the vortex tubes and forms double-peak structures centered on the
tubes. This difference of structure implies differences in intermittency
between enstrophy and dissipation and has been linked to a difference between
the empirical scaling exponents of longitudinal and transverse structure
functions recently observed in experimental measurements \cite{2,3,4,5} and
DNS \cite{6,7,8}.

Siggia \cite{9}, Kerr \cite{10}, and Meneveau et al. \cite{11} have noted
that enstrophy is more intermittent than dissipation. More recently, Chen et
al. \cite{6} used DNS data to calculate empirical scalings of enstrophy and
dissipation. They found that the exponents for locally-averaged enstrophy were
significantly smaller than those for locally-averaged dissipation and related
the differences to an elaboration of Kolmogorov's refined similarity
hypothesis.

Traditional experimental measurements typically invoke Taylor's hypothesis
and the calculation of inertial-range scaling exponents can be contaminated by
large-scale shear effects. DNS using the largest available computers can reach
only moderate Reynolds, at which the inertial range is narrow. These
difficulties accentuate the need for theoretical understanding of
intermittency phenomena and their role in scaling.

Distributions of idealized vortices have been used by a number of authors to
model turbulence statistics. The hope here is that the vortex distribution can
characterize both the essential physics of the small-scale structures and the
observed vorticity statistics, thereby leading to an acceptable description of
overall turbulence properties. Idealized vortices that have been studied
include the Hill vortex \cite{12}, simple vortex filaments \cite{13}, the
Burgers vortex \cite{14}, and spiral vortices \cite{14}. A distribution of
modified Burgers vortices will be used in the present paper to model scaling
exponents over finite ranges. Before presenting detailed results for the
Burgers vortex, we shall discuss the underlying question of whether
cylindrical vortex structures can support different asymptotic scaling
exponents for enstrophy and dissipation.

Moments of locally averaged enstropy and dissipation densities in a velocity
field with isotropic overall statistics may be defined by
\begin{equation}
\Omega_n(\ell) = \left\langle [\Omega]_\ell^n\right\rangle , \quad
\epsilon_n(\ell) = \left\langle [\epsilon]_\ell^n\right\rangle ,
\label{1}\end{equation}
where $\left\langle \ \right\rangle $ denotes ensemble average over the
isotropic statistics and $[\ ]_\ell$ denotes space average over a region of
characteristic linear dimension $\ell$. Joint powerlaw scaling with enstrophy
exponents $\zeta_n$ and dissipation exponents $\xi_n$ exists if there is a
range of $\ell$ in which
\begin{equation}
\Omega_n(\ell) \propto (L/\ell)^{\zeta_n}, \quad \epsilon_n(\ell) \propto
(L/\ell)^{\xi_n},
\label{2}\end{equation}
Here $L$ is a macroscale that marks the bottom of the scaling range. The
space averaging $[\ ]_\ell$ smears spots of intense excitation. Therefore
$\zeta_n$ and $\xi_n$ are expected to be positive. Let $\ell_d$ mark the top
of the scaling range, which we assume is also the beginning of the dissipation
range.

Suppose that $0 < \xi_n < \zeta_n$ and that the length of the scaling range
in decades becomes infinite in the limit $Re\to\infty$ where $Re$ is a
turbulence Reynolds number. Then $\Omega_n(\ell_d)/\epsilon_n(\ell_d) \to
\infty$ as $Re \to \infty$. A corollary is that the ratio of single-point
averages becomes infinite in the limit: $\left\langle \Omega^n\right\rangle
/\left\langle \epsilon^n\right\rangle  \to \infty$. We assume here that
$\left\langle \epsilon^n\right\rangle /\epsilon_n(\ell_d)$ does not become
infinitely larger than $\left\langle \Omega^n\right\rangle /\Omega_n(\ell_d)$.
Now we can ask what kind of vortex structure, if any, can support this
behavior. If $\left\langle \Omega^n\right\rangle /\left\langle
\epsilon^n\right\rangle $ goes infinite, where the averages are over the
entire field, then there must be at least one vortex structure in the field
for which the ratio of averages over that single structure goes infinite.

A cylindrical vortex is characterized by the azimuthal velocity
$v_\theta(r)$, where $r$ is distance from the vortex axis. The enstrophy
density and dissipation density (normalized by $\nu$) are
\begin{equation}
\Omega(r) = \left({dv_\theta\over dr} + {v_\theta\over r}\right)^2, \quad
\epsilon(r) = \left({dv_\theta\over dr} - {v_\theta\over r}\right)^2.
\label{3}\end{equation}
The enstrophy and dissipation per unit length of vortex are
$2\pi\int_0^\infty \Omega(r)r\,dr$ and $2\pi\int_0^\infty \epsilon(r)r\,dr$,
respectively.
The moments
\begin{equation}
\Omega_n = 2\pi\int_0^\infty [\Omega(r)]^nr\,dr, \quad \epsilon_n =
2\pi\int_0^\infty [\epsilon(r)]^nr\,dr
\label{4}\end{equation}
describe the distribution of enstrophy and dissipation densities in the
single vortex structure. By (\ref{3}), $\Omega_1 = \epsilon_1$ if $v_\theta(0)
= v_\theta(\infty) = 0$

Consider first the Rankin vortex of radius $r_0$ for which $v_\theta \propto
r$ $(r < r_0)$ and $v_\theta \propto 1/r$ $(r > r_0)$. Here the vorticity is
confined to a rigidly rotating core and all the dissipation lies outside the
core. The ratios $R_n = \Omega_n/\epsilon_n$ calculated from (\ref{3}) are
$R_n = 2n-1$.

For the Burgers vortex of radius $r_b$,
\begin{equation}
v_\theta(r) = {\Gamma\over2\pi}{1-\exp(-r^2/r_b^2)\over r},
\label{5}\end{equation}
where $\Gamma$ is the total circulation. By (\ref{3}), $\Omega(r) \propto
\exp(-2r^2/r_b^2)$, and the vorticity and dissipation now overlap.
Eq.~(\ref{3}) yields $R_2 \approx 10.65$, $R_3 \approx 104.07$, $R_4 \approx
1040.02$. To a rough approximation, $R_n \sim 10^{n-1}$.

The cylindrical vorticity distribution that maximizes $R_2$ can be found by
solving the associated variational problem, with $\int_0^\infty\omega(r)r\,dr$
and $\Omega_1$ held constant. The result confirms what can be guessed by
inspection of (\ref{3}): The maximizing distribution is the limit
$r_0\to\infty$ of
\[
v_\theta(r) \propto r\ \ (r < r_0), \quad v_\theta(r) \propto r^\alpha\ \
(r_0 < r < r_1),
\nonumber \]
\begin{equation}
v_\theta(r) \propto 1/r\ \ (r > r_1),
\label{6}\end{equation}
with $\alpha=1/2$. In this case, $R_n=(n-1)9^n$ for $n > 1$. The essential
facts here are first, if $v_\theta \propto r^\alpha$, then
$\Omega(r)/\epsilon(r)$ grows as $\alpha \to 1$; second, $\alpha=1/2$ is the
largest $\alpha$ for which $\int_{r_0}^{r_1}[\Omega(r)]^2r\,dr$ diverges as
$r_0\to0$, thereby making the relative contribution to $\epsilon_2$ from the
$v_\theta \propto 1/r$ region negligible in the limit. The maximum $R_n$ for
$n>2$ also are finite; they are maximized at $\alpha$ values that increase
with $n$.

It follows that differing asymptotic scaling exponents for enstrophy and
dissipation cannot be supported by cylindrical vortices. The vortex profile
that maximizes $R_2$ is one in which $\omega(r)$ is highly diffuse and the
dominant contributions to $[\Omega(r)]^2$ and $[\epsilon(r)]^2$ have the same
$r$ dependence. This refutes the intuition that the maximizing distribution is
one in which a compact vortex core is surrounded by diffuse dissipation.

Similar arguments establish that differing asymptotic scaling exponents for
longitudinal and transverse structure functions \cite{15} cannot be supported
by cylindrical vortices.

If the cylindrical vortex is replaced by a plane vortex layer, it is trivial
that all $R_n = 1$. This suggests that the cylindrical vortex may be the form
that maximizes the $R_n$. In any event, we conjecture that the $R_n$ are
finite whatever the shape of the finite volume that is filled with the intense
vorticity.

The Navier-Stokes (NS) equation so far has not been invoked. Highly diffuse
cylindrical vortices are not expected to survive under the NS equation; the
value $R_2=10.65$ for the Burgers vortex plausibly is closer to the maximum
$R_2$ attainable under NS.

In the remainder of this paper, a distribution of modified Burgers vortices
will be constructed in order to demonstrate that different finite-range
scaling exponents for enstrophy and dissipation can be realized despite the
fact that the infinite-range exponents must be equal. Data from experiments
and DNS are invoked in the construction, but the model is essentially
artificial and we do not offer it as a realistic representation of the vortex
structure of actual turbulence.

Exponential-like distributions of vortex core radii have been observed in DNS
by Jimen\'ez et al. \cite{19} and in an experiment by Tabeling et al.
\cite{21}. We introduce a distribution of core sizes into our model by fitting
the high-Reynolds number data in \cite{21} to a stretched exponential, thereby
obtaining the radius pdf
\begin{equation}
P(r_b) \propto \eta^{-3}r_b^2\exp[-(r_b/\eta)^{0.7}],
\label{7}\end{equation}
where $\eta$ is an assumed Kolmogorov dissipation length \cite{22}. Of course
the experimentally observed vortex cores are not Burgers vortices.

The Burgers vortex (\ref{5}) is in equilibrium under the NS equation if it
lies within a uniform background strain field with radial and axial strain
components $-a$ and $a$, such that $r_b=(2\nu/a)^{1/2}$. Here $\nu$ is
kinematic viscosity. This background field has no vorticity but it does
contribute to the total dissipation density, which becomes
\begin{equation}
\epsilon(r) = 12a^2 + \left({dv_\theta\over dr} - {v_\theta\over r}\right)^2.
\label{8}\end{equation}
We consider a distribution of isolated Burgers vortices, each in its own
background field. The ensemble average $\left\langle \ \right\rangle $ is over
the radius pdf (\ref{7}). The local averaging $[\ ]_\ell$ invoked in (\ref{1})
is realized as the average over a cylinder of radius $\ell$ and length
$2\ell$, centered on on a point ${\bf x}$ in the space surrounding a vortex
and aligned with the axis of the vortex, followed by an average over ${\bf
x}$. In order to have finite total dissipation per unit core length according
to (\ref{8}), the integration over ${\bf x}$ is extended only to a finite
distance $r_{max}$ from the vortex axis. This truncation is in lieu of taking
account of vortex interactions and making realistic changes in the straining
field far from the vortex core.

\bigskip
\psfig{file=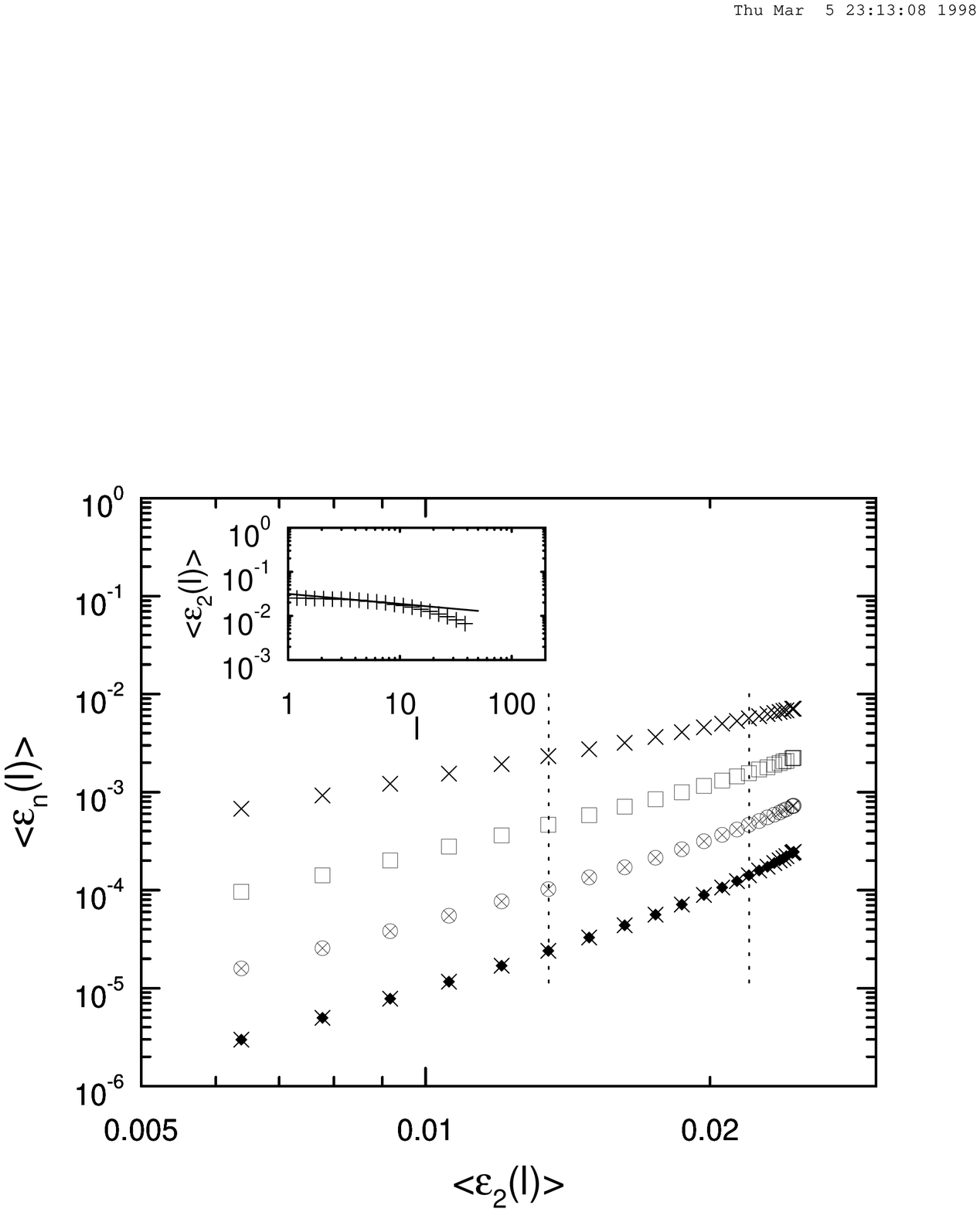,width=240pt}
\noindent
{\small FIG.~1. Parametric plot of $\epsilon_n(\ell)$ as a function of
$\epsilon_2(\ell)$ for $n = 3,4,5$ and $6$ (from top to bottom). Finite-range
exponents used in Figs.~2 and 5 are best fits over the region between the two
dashed lines. The inset shows $\epsilon_2(\ell)$ versus $\ell$.}
\bigskip

In the numerical study reported here, the space integrations needed to
evaluate the $\Omega_n(\ell)$ and $\epsilon_n(\ell)$ were performed by a
second-order trapezoidal scheme whose accuracy was verified. We have taken
$r_{max} = 10\eta$ and have fixed the vortex Reynolds number at $R_\Gamma
\equiv \Gamma/\nu = 1293$. These values, together with (\ref{7}),
(non-uniquely) make the inhomogeneous model satisfy the homogeneity relations
$\Omega_1(\ell) = \epsilon_1(\ell)$. By (\ref{5}), a change in the value of
$R_\Gamma$ does not affect the enstrophy scaling, but by (\ref{8}) it does
affect the relative scalings of enstrophy and dissipation.

\bigskip
\psfig{file=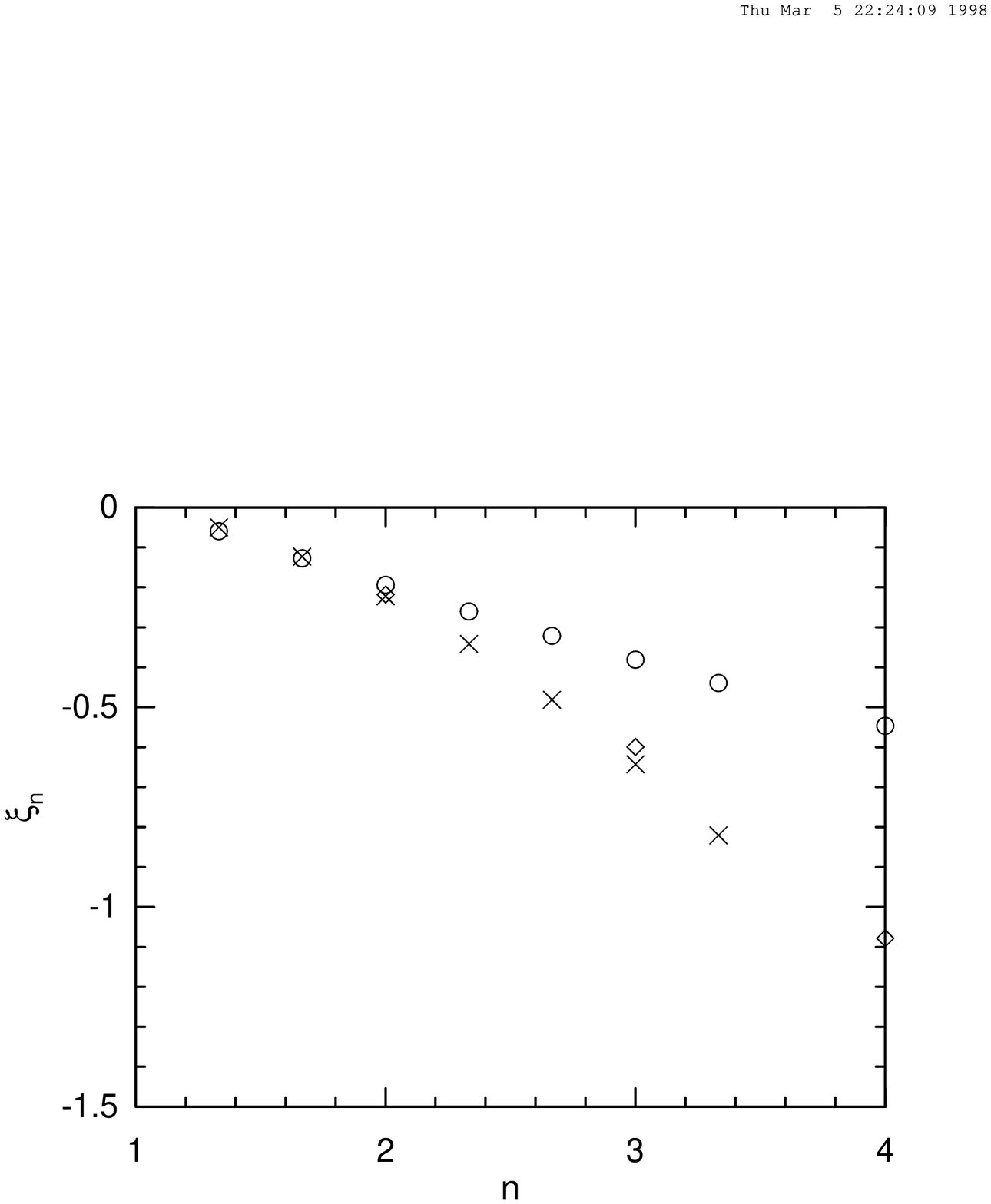,width=220pt}
\noindent
{\small FIG.~2. The scaling exponents $\xi_n$ of the locally averaged
dissipation as a function of $n$: The circles represent our Burgers vortex
model, the diamonds are data from an experiment by Sreenivasan et al.
\cite{23} and the crosses are DNS data \cite{6}.}
\bigskip

Fig. 1 shows the $\epsilon_n(\ell)$ plotted against $\epsilon_2(\ell)$ for
$n=3,4,5$. A narrow powerlaw range can be identified. In the inset, we plot
$\epsilon_2(\ell)$ against $\ell$. Local scaling exponents for dissipation and
enstrophy are calculated as $\xi_n = d\ln\,\epsilon_n(\ell)/d\ln\,\ell$ and
$\zeta_n = d\ln\,\Omega_n(\ell)/d\ln\,\ell$. The data used for Fig. 2 yield
$\xi_2 = 0.195 \pm 0.0747$, which is not far from the value $2/9$ given by
some simple scaling models or the experimental value $0.25 \pm 0.05$ \cite{23}.

Fig. 2 shows $\xi_n$ versus $n$. For $n \le 2$, the present model agrees well
with the DNS and experimental data. For $n > 2$, the exponents from the
present model decrease too slowly, indicating an unrealistic rate of increase
of intermittency with decrease of scale.

\bigskip
\psfig{file=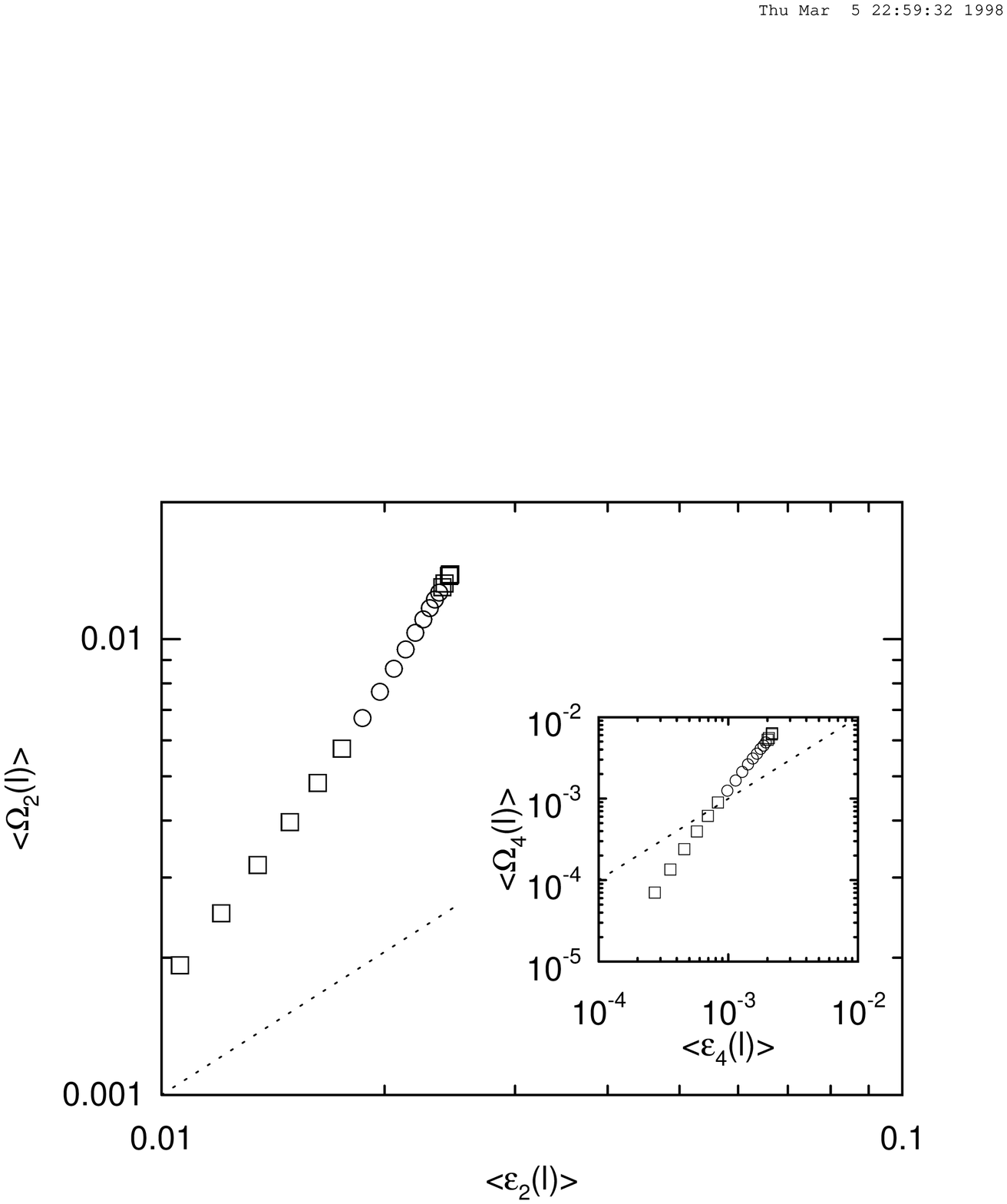,width=220pt}
\noindent
\vskip 5pt
{\small FIG.~3. Parametric plot of $\Omega_2(\ell)$ as a function of
$\epsilon_2(\ell)$. The circles denote the part of the plot for which $\ell$
lies between the dashed lines of Fig.~1. The dotted line corresponds to
$\Omega_2(\ell) = \epsilon_2(\ell)$. The inset is a similar plot of
$\Omega_4(\ell)$ and $\epsilon_4(\ell)$.}
\bigskip

\bigskip
\psfig{file=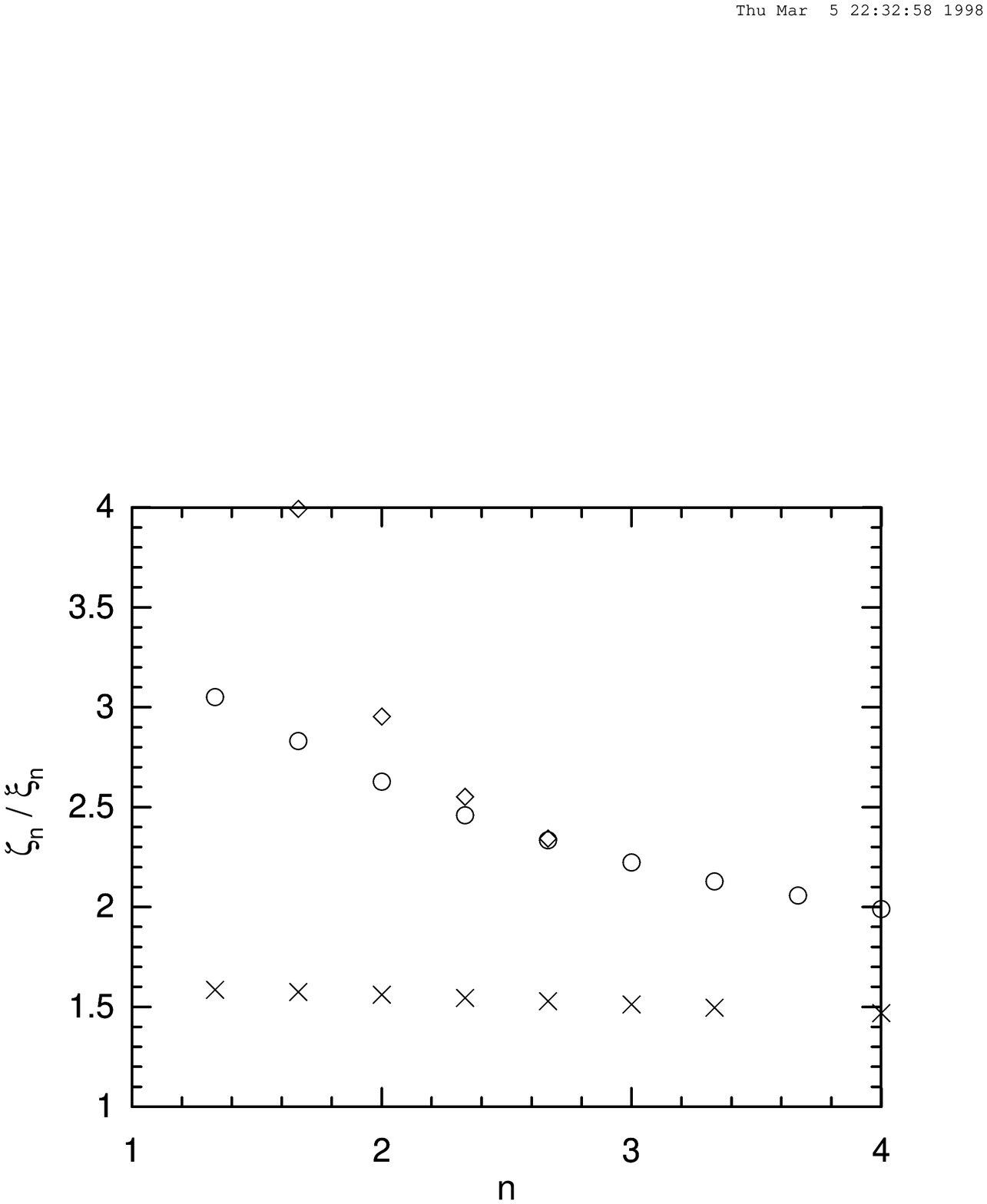,width=220pt}
\noindent
{\small FIG.~4. Ratio of enstrophy and dissipation scaling exponents
$\zeta_n/\xi_n$ as a function of $n$. The circles represent the present
Burgers' vortex model, the diamonds are data from an experiment by Antonia et
al \cite{5}, and the crosses are DNS data \cite{6}.}
\bigskip

In Fig. 3, $\Omega_2(\ell)$ and $\Omega_4(\ell)$ are plotted against
$\epsilon_2(\ell)$ and $\epsilon_4(\ell)$, respectively. Fig.~4 shows the
ratio $\zeta_n/\xi_n$ as a function of $n$ for $4/3 \le n \le 4$. Data are
presented for experiment \cite{5}, DNS \cite{6}, and the present model. The
DNS was computed at mesh resolution $512^3$. The DNS ratio values in Fig. 4
decrease very slowly with increase of $n$. For the experimental data, the
ratio values were calculated from measurements of longitudinal and transverse
velocity structure functions by invoking refined similarity hypotheses for
longitudinal velocity increments \cite{24} and transverse velocity increments
\cite{6}. Experiment, DNS, and model all yield $\zeta_n/\xi_n > 1$.

We reach two principal conclusions. First, if the support of intense
vorticity lies in cylindrical vortices, then differing asymptotic scaling
exponents for locally-averaged enstrophy and dissipation are impossible. We do
not see how this conclusion can be changed if the support lies in vortex
structures of non-cylindrical shape. Second, models built on cylindrical
vortices can yield differing finite-range scaling exponents
\cite{25}. It is plausible that compact cylindrical vortices do 
mediate the entrophy intermittency
measured at small scales. Moffatt et al \cite{18} found, from an
analysis of cylindrical vortices at a uniformly distributed angle to a
constant strain field, that the vortex cores contained 63.3\% of the
total enstrophy and only 1.3\% of the total enstrophy.

The Burgers vortex model presented here is arbitrary in a number of respects.
Nevertheless, it is plausible that the filamentary vortex cores of actual
turbulence are associated with moment ratios $R_n$ that have orders of
magnitude similar to those of the Burgers vortex. In this event, different
best-fit scaling exponents for enstrophy and dissipation can be induced over
substantial finite ranges of scales.

We thank H. Chen, M. Nelkin, K. R. Sreenivasan, Stefan Thomas,
E. Titi and J. Z. Wu for useful discussions.

\end{multicols}

\end{document}